\documentclass[prl,preprint]{revtex4-1}

\usepackage{amssymb}
\usepackage{amsmath}
\usepackage{amsfonts}
\usepackage{graphicx}

\newcommand{\beq}{\begin{eqnarray}}
\newcommand{\eeq}{\end{eqnarray}}

\begin{document}

\title{A comment on ``Interlayer interactions in graphites''
[Chen et al., Sci. Rep. 3, 3046 (2013)]}
\author{Tim Gould}%
\affiliation{Queensland Micro and Nano Technology Centre, Griffith
  University, Nathan, Queensland 4111, Australia}
%\author{John F. Dobson}%
%\affiliation{Queensland Micro and Nano Technology Centre, Griffith
%  University, Nathan, Queensland 4111, Australia}
\author{Tom\'a\v{s} Bu\v{c}ko}\affiliation{Comenius University in Bratislava,
Faculty of Natural Sciences, Slovakia, Theoretical Chemistry}
\author{S\'ebastien Leb\`egue}%
\affiliation{Laboratoire de Cristallographie, R\'esonance Magn\'etique
et Mod\'elisations (CRM2, UMR CNRS 7036)
Institut Jean Barriol, Universit\'e de Lorraine, BP 239,
Boulevard des Aiguillettes 54506 Vandoeuvre-l\`es-Nancy,France}
\begin{abstract}
Determining the material properties of layered systems like graphite
and bigraphene from \emph{ab initio} calculations is very difficult.
This is mostly due to the complex van der Waals forces
which help bind the layers. Recently, Chen~\emph{et al.}\cite{Chen2013}
reported a novel approach for extracting geometry dependent energetic
properties of general, layered graphitic systems from
periodic graphite calculations on AA, AB and ABC graphite.
Unfortunately, their analysis suffered from a number of
technical and theoretical flaws which make their results
unreliable for predicting energetic properties.
We propose that their conclusions in this regard
should be reassessed, or reanalysed
using more appropriate van der Waals
theory\cite{Gould2008,Gould2009,Lebegue2010,Gould2013-Cones}.
\end{abstract}
\maketitle

To determine graphitic material properties,
Chen~\emph{et al.}\cite{Chen2013} (CTPDC) utilise a number
of recent developments
in \emph{ab initio} theory in combination with a M\"obius inversion
method. While we believe their use of M\"obius inversion is
scientifically justified, at least within the limits of additive
van der Waals theory\cite{Gould2009,Dobson2012-JPCM,Gobre2013},
the \emph{ab initio} and analytic inputs
used are less well justified and will make their energetic predictions
unreliable. Indeed their application of Grimme's
DFT-D2 method\cite{Grimme2006}, and their fitting and
theoretical conclusions all seem to have fundamental flaws. Without
good theoretical input, the M\"obius inversion method cannot be
expected to produce appropriate quantitative conclusions, and
we show that this is indeed the case.

\begin{figure}[a]
\caption{Energetic curves from ACFD-RPA\cite{Lebegue2010},
compared with those reported by Chen \emph{et al.}\cite{Chen2013},
Gould, Simpkins and Dobson\cite{Gould2008} and
Gould, Leb\`egue and Dobson\cite{Gould2013-Model}.
Results are included for stretched
graphite (lower group of curves) and bigraphene (higher curves).
\label{fig:En}}
\includegraphics[width=1.0\linewidth]{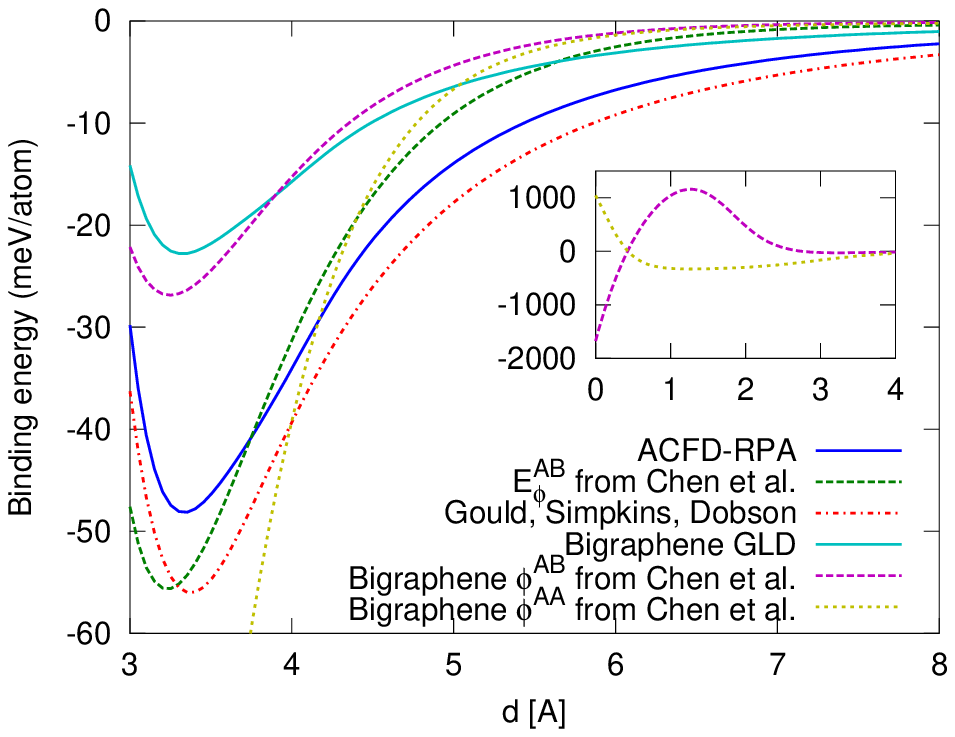}
\end{figure}
In their analysis\cite{Chen2013}, CTPDC appear to discard the
well-known asymptotic van der Waals power law $E=-C_pd^{-p}$
(relating the potential energy $E$ to the interlayer
distance $d$) completely from the 
Rahaman-Stillinger-Lemberg (RSL2) fitting functions
($\phi^{AB}$ and $\phi^{AA}$, their Equation 1) they use
for their M\"obius inversion.
Instead they replace it by a faster-decaying function that asymptotes
to an inverse exponential $E=-Ke^{-\kappa d}$. This disregards
the long-range component of the van der Waals forces and
makes extrapolation from one geometry to another problematic.
It leads to qualitative and quantitative discrepancies between
their results and theories that do include van der Waals forces.

\begin{table}
\caption{Geometrically determined differences in the
potential energy minima of AB graphite in meV/Atom
(for exfoliation: meV/Surface Atom).
\label{tab:Geom}}
\begin{ruledtabular}\begin{tabular}{lrrrr}
& Chen~\emph{et. al}\cite{Chen2013}
& Gould~\emph{et. al}\cite{Gould2013-Model}
& Bj\"orkman~\emph{et. al}\cite{Bjorkman2012}
\\
$E_{\text{Bigraphene}}-E_{\text{Graphite}}$ & 2.0 & 4.1 & --
\\
$E_{\text{Bigraphene}}-E_{\text{Exfoliation}}$ & 2.0 & 6.5 & 6.5
\end{tabular}\end{ruledtabular}
\end{table}
As a result, their model fails to appropriately treat energy
differences between different  geometrical arrangements, such as the
difference in the per atom energy at equilibrium between stretched
graphite and bigraphene.
We consider three cases: stretched bulk graphite (where an infinite
number of layers are separated equally to $d$ - with energy
given by $E_{\phi}^{AB}$ in Eq.~(2) of CTPDC),
bigraphene (where two layers are separated to $d$ -
given by $\phi^{AB}$ in Eq.~(1) of CTPDC)
and exfoliation (where one layer is separated to a distance $d$
from a bulk graphite surface - given by $E_{\text{exf}}$ in Eq.~(3) of CTPDC).
In Table~\ref{tab:Geom} we show
the energy gained by a bigraphene arrangement compared to
other arrangements.
Results using CTPDC's formulae show a significant difference from
previous predictions, a direct result of neglecting the
van der Waals power laws that contribute energy terms
beyond the nearest neighbour layers. Notably, the energetics of
exfoliation and graphite stretching are found to be identical
in their analysis, as their exponential terms cannot introduce
more than second-nearest neighbour contributions.

The poor energetics can be seen most prominently in the case of AA
graphite, where insertion of the AA parameters from Table 2
into Equation 1 of CTPDC\cite{Chen2013} gives a potential well for
$E_{\text{graph-AA}}=E^{AA}_{\phi}$ (from Equation 2 of their work)
with a depth of 13800meV/Atom located at $d_0=0.076$\AA.
Most other theories find an AA lattice spacing of
approximately 3.6\AA\ (see Table 1 of CTPDC).
This is almost certainly a result of neglecting
the van der Waals power law decay terms from
the AA bigraphene potential $\phi^{AA}$,
leading to a well of depth 328meV/Atom located at $d_0=1.29$\AA\ 
(see inset of Figure~\ref{fig:En}).
This is clearly an unphysical result, and makes portability
of the model to new geometries highly dubious.

The neglect of the vdW power law also means that their fit only applies
to energies very near contact, and thus cannot be used to predict
forces in the important intermediate and outer parts of the binding
curve  (for example the peak force occurs at $d\approx 3.7$\AA).
This is demonstrated in Figure~\ref{fig:En}
where their energy results
for bulk graphite $E^{AB}$ and bigraphene $\phi^{AA}$ and $\phi^{AB}$
are compared with theories\cite{Gould2008,Lebegue2010,Gould2013-Model}
that contain appropriate van der Waals power laws.
Here the high-level ACFD-RPA\cite{Lebegue2010}
is used as a benchmark for AB graphite
while the AB bigraphene prediction of Gould, Leb\`egue and
Dobson\cite{Gould2013-Model}
(GLD), which is guaranteed to reproduce numerous theoretical
and experimental properties of graphitic systems,  is used as a
benchmark for bigraphene in the absence of highly accurate
ACFD-RPA results. Apart from the near-contact region of the
AB case, their function fit gives substantially quantitatively
and qualitatively different results to previous theory, and this
discrepancy comes entirely from their neglect of an appropriate
van der Waals power law.

In addition to this very fundamental flaw, the paper also contains
other misinterpretations of previous theory. Firstly,
in their Table I, CTPDC\cite{Chen2013} report results
obtained using Grimme's D2 correction\cite{Grimme2006} on top
of either LDA or GGA-PBE. As detailed in the original paper\cite{Grimme2006},
Grimme's D2 correction is functional dependant via a factor that adjusts
the correction to the functional used for the DFT calculations. This
factor is obtained by least-squares optimization with respect to
a training set of systems in order to reproduce at best the interaction
energies and provided\cite{Grimme2006} for various functionals but not
for the LDA. Therefore in order to correct LDA, one should in principle
conduct the same procedure to obtain the corresponding factor, a fact
which is not mentioned by CDPBC\cite{Chen2013}.
It is likely that the authors have used the default parameter provided
by the implementation of Grimme's D2 method\cite{tomaspbed2}
in the VASP code\cite{kresse_efficiency_1996,kresse_ultrasoft_1999}
which aims at correcting the GGA-PBE functional, but not the LDA,
and therefore the LDA/DFT-D2 results in Table I of CTPBC
are doubtful. Beside this point, it is generally not advisable to correct
results obtained with LDA since this functional, albeit being purely
local by construction, gives sometimes a fictitious attraction between
van der Waals interacting systems which is even more difficult to
correct than a repulsive (or weakly attractive) functional such
as GGA-PBE.

\begin{figure}[a]
\caption{vdW behaviour as reported by Chen \emph{et al.}\cite{Chen2013},
and Gould, Simpkins and Dobson\cite{Gould2008}. Power law fits to
the results of GSD are included for comparison.
\label{fig:vdW}}
\includegraphics[width=1.0\linewidth]{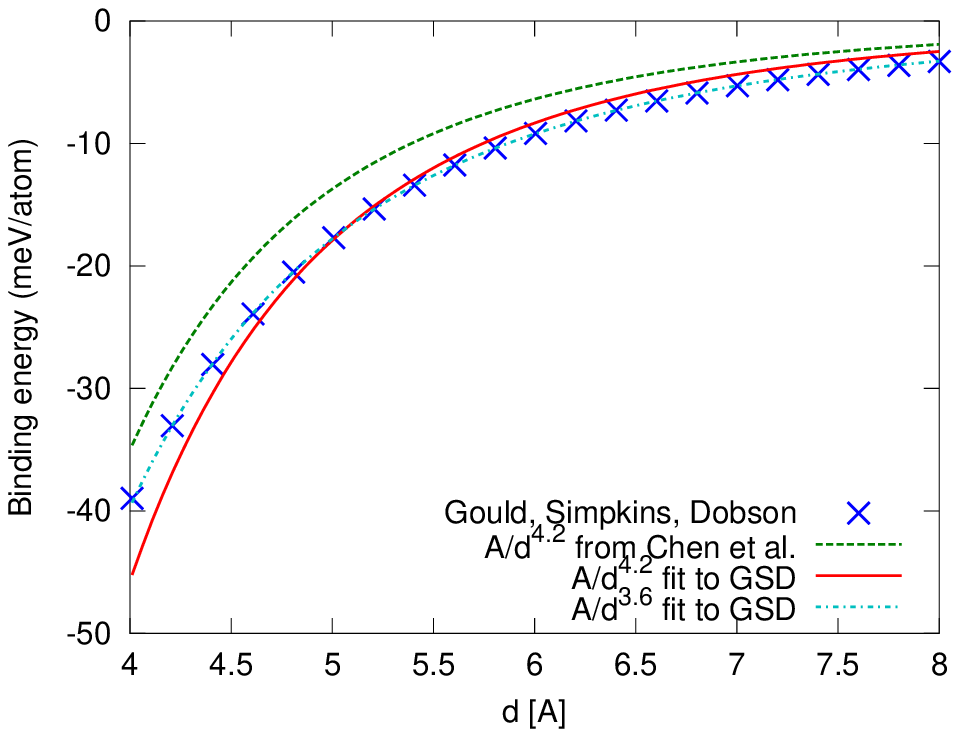}
\end{figure}
Secondly, CTPDC then justify their use of Grimme's functional
by comparing the energetics of stretched graphite with previous
random-phase approximation (RPA) theory studies to show that they
match the effective power law exponent in a given interlayer
distance range.
Here they make a fit $y=A/d^{4.2}+C$ and state that this agrees with
RPA results by Gould, Simpkins and Dobson\cite{Gould2008} (GSD) and
Chakarova-K\"ack~\emph{et al.}\cite{Chakarova2006}. However, using
the energy curve from GSD
yields $y=A'/d^{3.6}+C'$ in the same region,
more than half a power of $d$ higher and approaching
the asymptotic $y=A''/d^3+C''$ found
in the RPA\cite{Dobson2006,Gould2008,Lebegue2010,Gould2013-Cones}.
The value of 4.2 for the coefficient given in
Ref.~\onlinecite{Lebegue2010} was found on a range of
3-9\AA, including the region of substantial overlap and should thus
not be considered asymptotic.
The different asymptotic behaviours are illustrated in
Figure~\ref{fig:vdW} where we show the results from CTPDC
compared with those of GSD, and power law fits to GSD.
While their calculations may agree better with Ref.~\cite{Chakarova2006},
we note that it does not report
RPA results at all, at least in the conventional
``ACFD-RPA''\cite{Eshuis2012} sense.

Altogether, the neglect of appropriate van der Waals effects
by Chen~\emph{et al.} means that their energy results are likely to
be substantially erroneous, even with good inputs.
At best we expect their method to give a crude estimate of the
energy of different geometrical arrangements of graphite layers
(like exfoliation or cleavage), and even then only in the immediate
contact region.

\section*{Acknowledgements}
TG was funded by the Griffith University Areas of Strategic Interest
fund.

\section*{Author contributions}
The corresponding author is TG, who co-wrote the manuscript with
SL. TB and SL performed the computational analysis
of the \emph{ab initio} calculations. TG did analysis of the fitting
functions and semi-analytic work. All authors contributed to discussions.

%\bibliographystyle{naturemag}
%\bibliography{vanDerWaals,ACFDT,DFT,Wannier,Misc,Experiment,Hybrid,%
%QMGeneral,ISTLS,Frac,OEP,ChargeTransfer,Graphene,seb}
%\bibliography{comment}

\end{document}